\documentclass[12pt]{article}

\renewcommand{\baselinestretch}{1.1}

\newtheorem{theorem}{Theorem}

\newfont{\Mb}{msbm10}
\newcommand{\R}{\mbox{\Mb\symbol{82}}}

\newcommand{\rA}{\rightarrow}

\newcommand{\dA}{\downarrow}

\begin{document}

\title{\Large \bf  Classification of Energy Momentum
Tensors in \\ $n \geq 5$ Dimensional Space-times: a Review \\ }

\author{
M.J. Rebou\c{c}as\thanks{ {\tt reboucas@cbpf.br}  }, \ \ J.
Santos\thanks{ {\tt janilo@dfte.ufrn.br} } \ \ and \ \ A.F.F.
Teixeira \thanks{ {\tt teixeira@cbpf.br} }
\\
$~^{{\ast}\;{\ddagger}}~$Centro Brasileiro de Pesquisas F\'\i
sicas
\\        Departamento de Relatividade e Part\'\i culas \\
          Rua Dr.\ Xavier Sigaud 150 \\
          22290-180 Rio de Janeiro -- RJ, Brazil \\
\\
$^{\dagger}~$Universidade Federal do Rio Grande do Norte \\
             Departamento de F\'{\i}sica, Caixa Postal 1641 \\
             59072-970 Natal --   RN,  Brazil \vspace{2mm}  \\
        }

\date{\today}

\maketitle

\renewcommand{\baselinestretch}{0.8}
\begin{abstract}
Recent developments in string theory suggest that there might exist extra
spatial dimensions, which are not small nor compact. The framework of a
great number of brane cosmological models is that in which the matter
fields are confined on a brane-world embedded in five dimensions (the
bulk). Motivated by this we review the main results on the algebraic
classification of second order symmetric tensors in $5$-dimensional
space-times. All possible Segre types for a symmetric two-tensor are
found, and a set of canonical forms for each Segre type is obtained. A
limiting diagram for the Segre types of these symmetric tensors in $5$--D
is built. Two theorems which collect together some basic results on the
algebraic structure of second order symmetric tensors in $5$--D are
presented. We also show how one can obtain, by induction, the
classification and the canonical forms of a symmetric two-tensor on
$n$--dimensional ($n > 5 $) spaces from its classification in $5$--D
spaces, present the Segre types in $n$--D and the corresponding canonical
forms. This classification of symmetric two-tensors in any $n$--D spaces
and their canonical forms are important in the context of $n$-dimensional
brane-worlds context and also in the framework of $11$--D supergravity and
$10$--D superstrings.
\end{abstract}

\renewcommand{\baselinestretch}{1.3}

%**********************************************************************
\section{Introduction}    \label{intro}
\setcounter{equation}{0}
%**********************************************************************

Recent developments in string theory and its extension $M$-theory have
suggested a scenario in which the matter fields are confined on $3$--D
brane-world embedded in $1+3+d$ dimensions (the bulk). It is not necessary
for the $d$ extra spatial dimensions to be small and compact, which is a
fundamental departure from the standard Kaluza-Klein unification scheme.
Within the brane-world scenario only gravity and other exotic matter such
as the dilaton can propagate in the ($1+3+d\,$)--dimensional bulk.
Furthermore, in general the number $d$ of extra dimensions is not
necessarily equal to $1$. However, this general paradigm is often
simplified to a $5$--D context, where matter fields are restricted to a
$4$--D space-time while gravity acts in $5$--D~\cite{recent,old}. In this
{\em limited\/} $5$--D framework a great number work in \emph{brane-world
cosmology\/} has been done (see, for example,
\cite{BraxBruck2003,Langlois2003} and references therein). Motivated by
this scenario, in this article we present a review of our studies on the
algebraic classification, limits of Segre types and algebraic structures
of second order symmetric tensors (energy-momentum,
Einstein and Ricci tensors) in $5$--D and $n$--D space-times%
~\cite{5Djmp}~--~\cite{PRT}.

It is well known that a coordinate-invariant characterization of the
gravitational field in general relativity (GR) is given in terms of the
curvature tensor and a finite number of its covariant derivatives relative
to canonically chosen Lorentz frames~\cite{Cartan}~--~\cite{MacCSkea}. The
Riemann tensor itself is decomposable into three irreducible parts, namely
the Weyl tensor (denoted by $W_{abcd}$), the traceless Ricci tensor
($S_{ab} \equiv  R_{ab} - \frac{1}{4}\, R\, g_{ab}$) and the Ricci scalar
($R \equiv R_{ab}\, g^{ab}$). The algebraic classification of the Weyl
part of the Riemann tensor, known as Petrov classification, has played a
significant role in the study of various topics in general relativity.
However, for full classification of curvature tensor of nonvacuum
space-times one also has to consider the Ricci part of the curvature
tensor, which by virtue of Einstein's equations $G_{ab} = \kappa\, {\cal
T}_{ab} + \Lambda \,g_{ab}\,$ clearly has the same algebraic structure of
both the Einstein tensor $G_{ab} \equiv R_{ab} - \frac{1}{2}\, R\,
g_{ab}\,$ and the energy momentum tensor ${\cal T}_{ab}$.

The algebraic classification of a sym\-metr\-ic two-\-tensor locally
defined on a 4-\-di\-men\-sional Lorentzian manifold (such as the Ricci,
Einstein and energy momentum tensors) is known as Segre classification,
and has been discussed by several authors~\cite{Hall}. It is of interest
in at least three contexts: one is in understanding some purely
geometrical features of space-times~\cite{Churc}~--~\cite{CorHal}. The
second one is in classifying and interpreting matter field
distributions~\cite{Hall1}~--~\cite{SanRebTei}. The third is as part of
the procedure for checking whether apparently different space-times are in
fact locally the same up to coordinate
transformations~\cite{Cartan}~--~\cite{MacCSkea}. For examples of the use
of this invariant characterization in a class of G\"{o}del-type
space-times~\cite{RT1983} see~\cite{GTappl}.

Before presenting the scope of this article, for the sake of completeness,
let us briefly recall some historical points, and in the next two
paragraphs we mention, in passing,  other contexts where the results of
the present paper may also be of some interest.

The idea that the various interactions in nature might be unified by
enlarging the dimensionality of the space-time has a long history that
goes back to the works of Nordstr\"om, Kaluza and Klein
~\cite{Nordstrom1914,Klein1926}. Its earlier adherents were mainly
interested in extending general relativity, but a late increased interest
has been apparent in the particle physics community, especially among
those investigating super-symmetry.

An approach to the $5$--D non-compact Kaluza-Klein scenario, known as
space-time-matter (STM) theory, has also been discussed in a number of
papers (see, e.g., \cite{WessonLeon1992}~--~\cite{Wesson1999} and
references therein; and also~\cite{PRT},~\cite{RTa}~--~\cite{CRT} ).

In this paper we review our main results on the algebraic classification,
limits and their relation to Segre types in $5$--D, and algebraic
properties of second order symmetric tensors (energy momentum, Einstein
and Ricci) in $5$--D space-times~\cite{5Djmp}~--~\cite{PRT}. We note that
the proof of Theorem~\ref{eigentheo}, the consequent treatment of the
classification in Segre types and the corresponding canonical forms
presented here are new. We also show how one can obtain, by induction, the
classification and the canonical forms of a symmetric two-tensor on
$n$-dimensional ($n > 5$) spaces from its classification in
$5$-dimensional spaces. This classification of symmetric two-tensors in
any $n$-dimensional spaces and their canonical forms are important in the
context of $n$-dimensional brane-worlds context as well as in the
framework of $11$--D supergravity and $10$--D superstrings.

To make this paper as clear and selfcontained as possible, in the next
section we introduce the notation and the underlying mathematical setting,
which will be used throughout the article. In section~\ref{class} we
present a new proof of a theorem previously stated in~\cite{HRST} (see
Lemma~3.1). Using this theorem (Theorem~\ref{eigentheo}) and the
classification of an arbitrary symmetric two-tensors $S$ defined on a
$4$--D space-time, we derive a classification of any second order
symmetric tensors in $5$--D, and obtain a set of canonical forms for each
class. In section~\ref{Limits} we recall the concept of limits of
space-time, hereditary properties, and build a limiting diagram for the
Segre types in $5$--D. In section~\ref{class-n} we show how one can
obtain, by induction, the algebraic classification and the canonical forms
of symmetric two-tensors on a $n$-dimensional space (general arena of
brane-worlds theories) from their classification in $5$--D.  In section%
~\ref{FurtherRes} we collect together in two theorems some basic results 
regarding eigenvectors and invariant subspaces of $R^a_{\ b}$, which 
generalizec theorems on $4$--D space-times~\cite{Hall,Hall3}. In the last 
section we present some concluding remarks.

Although the Ricci and the energy momentum tensors are constantly referred
to, our results apply to any second order real symmetric tensor defined on
$5$--D (and sometimes $n$--D) Lorentzian manifolds.

%**********************************************************************
\section{Prerequisites}     \label{pre}
\setcounter{equation}{0}
%**********************************************************************

In this section we shall fix our general setting, define the
notation and briefly recall some important results required for
the remainder of this work.

The algebraic classification of the Ricci tensor at a
point $p \in M$ can be cast in terms of the eigenvalue problem
\begin{equation} \label{eigeneq}
(R^{a}_{\ b}-\lambda \, \delta^{a}_{b})\,V^{b}\, = 0 \;,
\end{equation}
where $\lambda$ is a scalar, $V^{b}$ is a vector and the mixed Ricci
tensor $R^{a}_{\ b}$ may be thought of as a linear operator $R: T_{p}(M)
\longrightarrow T_{p}(M)$. In this work $M$ is a real $5$-dimensional (or
$n$-dimensional, in section~\ref{class-n}) space-time manifold locally
endowed with a Lorentzian metric of signature ${\rm(}- + + + +{\rm)}\,$,
[or $\,{\rm(}- +\,  \cdots \, + {\rm)}$ in $n$--D case]  $\; T_{p}(M)$
denotes the tangent space to $M$ at a point $p \in M$ and Latin indices
range from $0$ to $4$, unless otherwise stated. Although the matrix
$R^a_{\ b}$ is real, the eigenvalues $\lambda$ and the eigenvectors $V^b$
are often complex. A mathematical procedure used to classify matrices in
such a case is to reduce them through similarity transformations to
canonical forms over the complex field. Among  the existing canonical
forms the Jordan canonical form (JCF) turns out to be the most appropriate
for a classification of $R^a_{\ b}$.

A  basic result in the theory of JCF~\cite{shilov} is that given
an $n$-square matrix $R$ over the complex field, there exist
nonsingular matrices $X$ such that
\begin{equation}
 \label{simtran}
    X^{-1}R\,X = J\;,
\end{equation}
where $J$, the JCF of $R$, is a block diagonal matrix, each block
being of the form
\begin{equation}   \label{jblock}
 J_{r}(\lambda_{k})=\left[
 \begin{array}{ccccc}
 \lambda_{k} &    1        &   0    & \cdots & 0 \\
      0      & \lambda_{k} &   1    & \cdots & 0 \\
             &             &        & \ddots &   \\
      0      &    0        &   0    & \cdots & 1 \\
      0      &    0        &   0    & \cdots & \lambda_{k}
  \end{array}
  \right].
\end{equation}
Here $r$ is the dimension of the block and $\lambda _{k}$ is the
$k$-th root of the characteristic equation  $\mbox{det}(R -
\lambda I)=0$. Hereafter  $R$ will be the real matrix formed with
the mixed components $R^{a}_{\ b}$ of the Ricci tensor.

A Jordan matrix $J$ is uniquely defined up to the ordering of
the Jordan blocks. Further, regardless of the dimension
of a specific Jordan block there is one and only one
independent eigenvector associated to it.

In the Jordan classification two square matrices are said to be
equivalent if similarity transformations exist such that they can
be brought to the same JCF. The JCF of a matrix gives explicitly
its eigenvalues and makes apparent the dimensions of the Jordan
blocks. However, for many purposes a somehow coarser
classification of a matrix is sufficient. In the Segre
classification, for example, the value of the roots of the
characteristic equation is not relevant --- only the dimension of
the Jordan blocks and the degeneracies of the eigenvalues matter.
The Segre type is a list $[r_1 r_2 \cdots r_m]$ of the dimensions
of the Jordan blocks. Equal eigenvalues in distinct blocks are
indicated by enclosing the corresponding digits inside round
brackets. Thus, for example, in the degenerated Segre type
$[(21)11]$ three out of the five eigenvalues are equal; there are
four linearly independent eigenvectors, two of which are
associated to Jordan blocks of dimensions 2 and 1, and the last
eigenvectors corresponds to two blocks of dimension 1.

In classifying symmetric tensors in a Lorentzian space two
refinements to the usual Segre notation are often used. Instead of
a digit to denote the dimension of a block with complex eigenvalue
a letter is used, and the digit corresponding to a time-like
eigenvector is separated from the others by a comma.

In this work we shall deal with two types of pentad of vectors,
namely the semi-null pentad basis $\{{\bf l},{\bf m},{\bf x},{\bf
y},{\bf z}\}$, whose non-vanishing inner products are only
\begin{equation}
l^{a}m_{a} = x^{a}x_{a} = y^{a}y_{a} = z^{a}z_{a} = 1 \;,
\label{inerp1}
\end{equation}
and the Lorentz pentad basis $\{{\bf t},{\bf w},{\bf x},{\bf
y},{\bf z} \}\,$, whose only non-zero inner products are
\begin{equation}
-t^a t_a = w^a w_a = x^a x_a = y^a y_a = z^a z_a = 1\;.
\label{inerp2}
\end{equation}

At a point $p \in M$ the most general decomposition of $R_{ab}$
in terms of a Lorentz basis for symmetric tensors at $p \in M$
is manifestly given by
\begin{eqnarray}
R_{ab} & = &
\sigma_{1}\,t_a t_b + \sigma_2\,w_a w_b   + \sigma_3\,x_a x_b +
\sigma_{4}\,y_a y_b + \sigma_{5}\,z_a z_b +
2\,\sigma_{6}\,t_{(a}w_{b)}                   \nonumber \\
&  &
+ \,2\,\sigma_{7}\,t_{(a}x_{b)} + 2\,\sigma_{8}\,t_{(a}y_{b)} +
 2\,\sigma_{9}\,t_{(a}z_{b)}  + 2\,\sigma_{10}\,w_{(a}x_{b)} +
 2\,\sigma_{11}\,w_{(a}y_{b)}                  \nonumber \\
&  & +\, 2\,\sigma_{12}\,w_{(a}z_{b)} +
2\,\sigma_{13}\,x_{(a}y_{b)} + 2\,\sigma_{14}\,x_{(a}z_{b)}+
2\,\sigma_{15}\,y_{(a}z_{b)} \;, \label{rabgen2}
\end{eqnarray}
where the coefficients $\sigma_1, \ldots ,\sigma_{15} \in \R $.

%**********************************************************************
\section{Classification in 5--D}     \label{class}
\setcounter{equation}{0}
%**********************************************************************

In this section we shall present the algebraic classification of symmetric
two-tensors defined on $5$--D space-times, which is based upon two
ingredients: one is the classification of second order symmetric tensors
in $4$--D, and  the other is a theorem stated below (proved
in~\cite{HRST}; see Lemma~3.1). We emphasize, however, the proof presented
here is completely new. \vspace{2mm}
\begin{theorem} \label{eigentheo}
Let $M$ be a real $5$-dimensional manifold endowed with a Lorentzian
metric $g$ of signature ${\rm(} - + + + +{\rm)}$. Let $R^a_{\ b}$ be the
mixed form of a second order symmetric tensor $R$ defined at any point $p
\in M$. Then $R^a_{\ b}$ has at least one real non-null eigenvector with
real eigenvalue.
\end{theorem}

{\bf Proof}. We initially  consider the cases when all eigenvalues
of $R$ are real. Suppose first that $R^a_{\ b}$ has a single
eigenvector. According to the above results and eq.\
(\ref{simtran}) it can be brought to a Jordan canonical form
$J^a_{\ b}$ with only one Jordan block, namely
\vspace{2mm}
\begin{equation}  \label{rab5}
 J^a_{\ b}\,=\,\left[
  \begin{array}{ccccc}
  \lambda &    1    &    0    &    0    &    0      \\
       0  & \lambda &    1    &    0    &    0      \\
       0  &    0    & \lambda &    1    &    0       \\
       0  &    0    &    0    & \lambda &    1      \\
       0  &    0    &    0    &    0    & \lambda
  \end{array}
  \right] \;, \\
\end{equation}
where $\lambda \in \R $. Clearly in this case the Segre type for
$R^a_{\ b}$ is $[5]$. The matricial equation (\ref{simtran})
implies that $R\,X=J\,X$. Using $J$ given by (\ref{rab5}) and
equating the corresponding columns of both sides of this equation
one obtains
\begin{eqnarray}
R\,{\bf X_1} & = & \lambda {\bf X_1}              \;,
\label{first} \\ R\,{\bf X_2} & = & \lambda {\bf X_2} + {\bf X_1}
\,, \label{second} \\ R\,{\bf X_3} & = & \lambda {\bf X_3} + {\bf
X_2} \;, \label{third} \\ R\,{\bf X_4} & = & \lambda {\bf X_4} +
{\bf X_3} \;, \label{fourth} \\ R\,{\bf X_5} & = & \lambda {\bf
X_5} + {\bf X_4} \;, \label{fifth}
\end{eqnarray}
where we have denoted by ${\bf X_A}$ (${\bf A} = 1, \cdots ,5$)
the column vectors of the matrix $X$. Clearly ${\bf X_1}$ is an
eigenvector of $R$. Since $R$ is a symmetric two-tensor, from
equations (\ref{first}) and ({\ref{second}) one finds that ${\bf
X_1}$ is a null vector. Thus, if $R^a_{\ b}$ admits a single
eigenvector it must be null. However, the Lorentzian character of
the metric $g$ on $M$, together with the symmetry of $R_{ab}\,$,
rule out this Jordan canonical form for $R^a_{\ b}$. Indeed, from
equations (\ref{first}), (\ref{third}) and (\ref{fourth}) one can
easily show that ${\bf X_1}.{\bf X_2} = {\bf X_1}.{\bf X_3} = 0$,
which together with eqs.\ (\ref{second}) and (\ref{third}) imply
${\bf X_2}.{\bf X_2} = 0$. Thus, ${\bf X_1}$ and ${\bf X_2}$ are
both null and orthogonal one to another. As the metric $g$ is
Lorentzian they must be proportional, hence $X$ is a singular
matrix. So, at a point $p \in M$ the Ricci tensor $R$ cannot be of
Segre type $[5]$, it cannot have a single eigenvector.

Suppose now that $R^a_{\ b}$ has two linearly independent
eigenvectors (${\bf k},{\bf n}$) and that they are both
null vectors. Let $\mu$ and $\nu$ be the associated
eigenvalues. Then
\begin{eqnarray}
R^{a}_{\ b}\, k^b &=& \mu \, k^{a} \;, \label{eigen1} \\
 R^{a}_{\ b}\, n^b &=& \nu \,  n^{a}  \;, \label{eigen2}
\end{eqnarray}
where $\mu, \nu \in \R $.
Since $R_{ab}$ is symmetric and (${\bf k},{\bf n}$) are
linearly independent, eqs.\ (\ref{eigen1}) and (\ref{eigen2})
imply that $\mu=\nu$. Thus, from eqs. (\ref{eigen1}) and
(\ref{eigen2}) the vector ${\bf v} = {\bf k} + {\bf n}$
is also an eigenvector of $R$ with real eigenvalue
$\mu$. Since ${\bf k}$ and ${\bf n}$ are linearly
independent null vectors it follows that
$v^a v_a = 2\,k^a n_a \not= 0$, so ${\bf v}$
is a non-null eigenvector with real eigenvalue.

In the cases where $R^a_{\ b}$ has more than two linearly
independent null eigenvectors one can always use two of them
to similarly construct one non-null eigenvector with real
eigenvalue.
Thus, when all eigenvalues of $R$ are real $R^a_{\ b}$ has
at least one non-null eigenvector with real eigenvalue.

The case when the Ricci tensor has complex eigenvalues can be
dealt with by using an approach borrowed from~\cite{Hall} as
follows. Suppose that $\alpha \pm i\beta $ are complex eigenvalues
of $R^{a}_{\ b}$ corresponding to  the eigenvectors ${\bf
V}_{\pm}= {\bf Y}\pm i{\bf Z}$, where $\alpha$ and $\beta \neq 0$
are real and ${\bf Y}, \, {\bf Z}$  are independent vectors
defined on $T_{p}(M)$. Since  $R_{ab}$ is symmetric and the
eigenvalues are different, the eigenvectors must be orthogonal and
hence equation ${\bf Y}.{\bf Y} + {\bf Z}.{\bf Z} = 0$ holds. It
follows that either one of the vectors ${\bf Y}$ or ${\bf Z}$ is
time-like and the other space-like or both are null and, since
$\beta \neq 0$, not collinear. Regardless of whether ${\bf Y}$ and
${\bf Z}$ are both null vectors or one time-like and the other
space-like, the real and the imaginary part of (\ref{eigeneq})
give
\begin{eqnarray}
R^{a}_{\ b}\,Y^{b} & = & \alpha Y^{a} - \beta Z^{a}\;,
\label{relc1}  \\ R^{a}_{\ b}\,Z^{b} & = & \beta Y^{a} + \alpha
Z^{a}\,.  \label{relc2}
\end{eqnarray}
Thus, the vectors ${\bf Y}$ and ${\bf Z}$ span a time-like
2-dimensional subspace  of $T_p(M)$ invariant under $R^{a}_{\
b}\,$. Besides, by a similar procedure to that used in%
~\cite{5Djmp} one can show that the $3$-dimensional space
orthogonal to this time-like 2-space is space-like, is also
invariant under $R^{a}_{\ b}$ and contains three orthogonal
eigenvectors of $R^{a}_{\ b}$ with real eigenvalues. Thus, when
$R^a_{\ b}$ has complex eigenvalues we again have at least one
non-null eigenvector (actually we have at least three) with real
eigenvalue. This completes the proof of Theorem~\ref{eigentheo}.

We shall now discuss the algebraic classification of the Ricci
tensor. Let ${\bf v}$ be the real non-null eigenvector of Theorem%
~\ref{eigentheo} and let $\eta \in \R $ be the corresponding eigenvalue.
Obviously the normalized vector ${\bf u}$ defined by
\begin{equation}   \label{vectoru}
u^a \,= \, \frac{v^a}{\sqrt{\epsilon \,v^a v_a}}
\qquad\mbox{with}\qquad
\epsilon \,\equiv  \, \mbox{sign} \: (v^a v_a) \,=\,
u^a u_a
\end{equation}
is also an eigenvector of $R^a_{\ b}$ associated to $\eta$.

If the vector ${\bf u}$ is time-like ($\epsilon = -1$), one can
choose it as the time-like  vector ${\bf t}$ of a Lorentz pentad
$\{{\bf{t}},{\bf{\tilde{w}}},
{\bf{\tilde{x}}},{\bf{\tilde{y}}},{\bf{\tilde{z}}}\}$. The fact
that ${\bf u} \equiv {\bf t}$ is an eigenvector of $R^a_{\ b}$ can
then be used to reduce the general decomposition (\ref{rabgen2})
to
\begin{eqnarray}    \label{rabt}
R_{ab} & = & -\eta\,t_a t_b +\sigma_2\,\tilde{w}_a\tilde{w}_b +
\sigma_3\,\tilde{x}_a\tilde{x}_b +
\sigma_{4}\,\tilde{y}_a\tilde{y}_b
+\sigma_{5}\,\tilde{z}_a\tilde{z}_b
+2\,\sigma_{10}\,\tilde{w}_{(a}\tilde{x}_{b)}  \nonumber \\ &  &
+\;2\,\sigma_{11}\,\tilde{w}_{(a}\tilde{y}_{b)}\,
+\,2\,\sigma_{12}\,\tilde{w}_{(a}\tilde{z}_{b)}
+\,2\,\sigma_{13}\,\tilde{x}_{(a}\tilde{y}_{b)}
+\,2\,\sigma_{14}\,\tilde{x}_{(a}\tilde{z}_{b)} \nonumber \\ &  &
+\;2\,\sigma_{15}\,\tilde{y}_{(a}\tilde{z}_{b)} \;,
\end{eqnarray}
where $\eta = - \sigma_1$. Using (\ref{inerp2}) and (\ref{rabt})
one finds that the mixed matrix $R^a_{\ b}$ takes the block
diagonal form
\begin{equation}
R^a_{\ b}\, =\, S^a_{\ b} - \,\eta\, t^a t_b\,. \label{rabt1}
\end{equation}
The first block is a $(4 \times 4)$ symmetric matrix acting on the
$4$--D space-like vector space orthogonal to the subspace $\cal U$
of $T_p(M)$ defined by ${\bf u}$. Hence it can be diagonalized by
spatial rotation of the basis vectors
(${\bf{\tilde{w}}},{\bf{\tilde{x}}},{\bf{\tilde{y}}},
{\bf{\tilde{z}}}$). The second block is 1-dimensional and acts on
the subspace $\cal U$. Thus, there exists a Lorentz pentad
%$\{{\bf t},{\bf w},{\bf x},{\bf y},{\bf z}\}$
relative to which $R^a_{\ b}$ takes a diagonal form
with real coef\/ficients. The Segre type for $R^a_{\ b}$ is
then $[1,1111]$ or one of its degeneracies.

If ${\bf u}$ is space-like ($\epsilon = 1$) one can choose it as
the space-like vector ${\bf z}$ of a  Lorentz pentad and using
eqs.\ (\ref{inerp2}) one similarly finds that $R^a_{\ b}$ takes
the block diagonal form
\begin{equation}
R^a_{\ b}\, =\, S^a_{\ b} + \,\eta\, z^a z_b\;, \label{rabz}
\end{equation}
where $\eta = \sigma_5 $. But now the $4$--D vector space orthogonal to
the space-like subspace of $T_p(M)$ defined by ${\bf u}$ is Lorentzian.
Then the mixed matrix $S^a_{\ b}$ effectively acts on a $4$--D Lorentzian
vector space and is not necessarily symmetric, it is not diagonalizable
in general. As $S_{ab}$ is obviously symmetric, from equation (\ref{rabz})
it follows that the algebraic classification of $R^a_{\ b}$ and a set of
canonical forms for $R_{ab}$ can be achieved from the classification of a
symmetric two-tensor $S$ on a $4$--D space-time.

Thus, using the known classification \cite{Hall} for $4$--D
space-times it follows that semi-null pentad bases can be
introduced at  $p \in M$ such that the possible Segre types and
the corresponding canonical forms for $R$ are given by
\begin{eqnarray}  \hspace{-1.0cm}
\mbox{\bf Segre type}& &\quad\qquad\qquad\mbox{\bf Canonical
form}\nonumber\\ {[}1,1111]  & R_{ab} = & 2\,\rho_1\,l_{(a}m_{b)}
+ \rho_2\,(l_{a}l_{b} + m_{a}m_{b}) + \rho_3\,x_{a}x_{b} +
\rho_4\,y_{a}y_{b} + \rho_5\,z_{a}z_{b} \;,   \label{rab11111}  \\
{[}2111]  & R_{ab} = & 2\,\rho_1\,l_{(a}m_{b)} \pm l_{a}l_{b} +
\rho_3\,x_{a}x_{b} + \rho_4\,y_{a}y_{b} + \rho_5\,z_{a}z_{b}\;,
                                                   \label{rab2111}  \\
{[}311]   & R_{ab} = & 2\,\rho_1\,l_{(a}m_{b)} + 2\,l_{(a}x_{b)} +
\rho_1\,x_{a}x_{b} + \rho_4\,y_{a}y_{b} + \rho_5\,z_{a}z_{b}\;,
                                               \label{rab311} \\
{[}z\,\bar{z}\,111]& R_{ab} = & 2\,\rho_1\,l_{(a}m_{b)} +
\rho_2\,(l_{a}l_{b} - m_{a}m_{b}) +\rho_3\,x_{a}x_{b} +
\rho_4\,y_{a}y_{b} + \rho_5\,z_{a}z_{b}\;,
                                                \label{rabzz111}
\end{eqnarray}
and the twenty-two degeneracies thereof, in agreement with Santos
{\em et al.\/} \cite{5Djmp}. Here $\rho_1, \cdots ,\rho_5 \in \R$
and $\rho_2\, \neq 0$ in (\ref{rabzz111}).

%**********************************************************************
\section{Limits and Segre Types in 5--D}  \label{Limits}
\setcounter{equation}{0}
%**********************************************************************

We shall use in this section the concept of limit of a space-time
introduced in reference~\cite{PaivaReboucasMacCallum1993}, wherein
by a limit of a space-time, broadly speaking, we mean a limit of a
family of space-times as some free parameters are taken to a
limit. For instance, in the one-parameter family of Schwarzschild
solutions each member is a Schwarzschild space-time with a
specific value for the mass parameter $m$.

In the study of limits of space-times it is worth noticing that
there are some properties that are inherited by all limits of a
family of space-times~\cite{Geroch1969}. These properties are
called hereditary. Thus, for example, a hereditary property
devised by Geroch can be stated as follows:
\begin{equation} \label{PropH0}
\begin{minipage}[t]{60ex}
{\bf Hereditary property:}\\ Let $T$ be a tensor or scalar field
built from the metric and its derivatives. If $T$ is zero for all
members of a family of space-times, it is zero for all limits of
this family.
\end{minipage}
\end{equation}

{}From this property we easily conclude that the vanishing of
either the Weyl or Ricci tensor or the curvature scalar are also
hereditary properties. In other words, limits of conformally flat
space-times are conformally flat, and that limits of Ricci flat
space-times are also vacuum solutions.

What can be said about the Petrov and Segre types of those tensors under
limiting processes? In general, the algebraic type of the Weyl tensor is
not a hereditary property under limiting processes. Nevertheless, to be at
least as specialized as the types in the Penrose specialization
diagram~\ref{PetrovEsp} for the Petrov classification is a hereditary
property~\cite{Geroch1969}.
\begin{figure}[h]
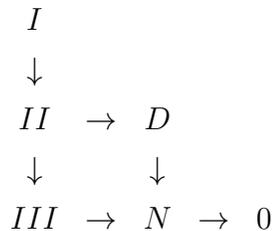
 \[
\begin{array}{ccccc}
I   &     &     &     &   \\ \dA &     &     &     &   \\ II  &
\rA & D   &     &   \\ \dA &     & \dA &     &   \\ III & \rA & N
& \rA & 0
\end{array}  \]
\caption{Limiting diagram for the Petrov types in $4$--D. }
\label{PetrovEsp}
\end{figure}

For simplicity, in the limiting diagrams in this work, we do not
draw arrows between types whenever a compound limit exists. Thus,
in figure~\ref{PetrovEsp}, e.g., the limits $I \rA II \rA D$ imply
that the limit $I \rA D$ is allowed.

In 1993 a coordinate-free technique for studying the limits of vacuum
space-times was developed and the limits of some well known vacuum
solutions were investigated~\cite{PaivaReboucasMacCallum1993}. In that
approach the Geroch limiting diagram for the Petrov classification was
extensively used. Five years later, to deal with limits on non-vacuum
solutions space-times in GR, Paiva {\em et
al.}~\cite{PaivaReboucasHallMacCallum1998} built a limiting diagram for
the Segre types in $4$--D. The main results of this paper is that the
Segre type of the Ricci tensor is not in general preserved under limiting
processes. However, under such processes the Segre types have to be at
least as specialized as the types in their diagram for the limits of the
Segre type they obtained (see figure~4 of
reference~\cite{PaivaReboucasHallMacCallum1998}).

The characteristic polynomial associated to the eigenvalue
problem~(\ref{eigeneq}), given by
\begin{equation}
\left| R^a_{\  b} - \lambda \delta^a_b \right| \;, \label{PC}
\end{equation}
is a polynomial of degree five in $\lambda$, and can be always
factorized over the complex field as
\begin{equation} \label{PoliCarac}
(\lambda - \lambda_1)^{d_1} \, (\lambda - \lambda_2)^{d_2} \,\,
\cdots \,\, (\lambda - \lambda_r)^{d_r} \;,
\end{equation}
where $\lambda_i$  ($i=1,2,\, \cdots \, ,r$) are the distinct
roots of the polynomial (eigenvalues), and $d_i$ the corresponding
degeneracies. Clearly $d_1+\,\cdots\,+d_r=5$. To indicate the
characteristic polynomial we shall introduce a new list \{$d_1\,
d_2\, \cdots \,d_r$\} of eigenvalues degeneracies, referred to as
the type of the characteristic polynomial.

The minimal polynomial can be introduced as follows. Let $P$ be a
monic matrix polynomial of degree $n$ in $R^a_{\ b}$, i.e.,
\begin{equation}
P = R^n + c_{n-1} \, R^{n-1} + c_{n-2} \, R^{n-2} + \, \cdots \, +
c_1\, R + c_0\,\delta \;,
\end{equation}
where $\delta$ is the identity matrix and $c_n$ are, in general,
complex numbers. The polynomial $P$ is said to be the minimal
polynomial of $R$ if it is the polynomial of lowest degree in $R$
such that $P=0$. It can be shown that the minimal (monic)
polynomial is unique and can be factorized as
\begin{equation} \label{PoliMin}
(R - \lambda_1 \delta)^{m_1} \, (R - \lambda_2 \delta)^{m_2} \,
\cdots \, (R - \lambda_r \delta)^{m_r} \;,
\end{equation}
where $m_i$  is the dimension of the Jordan block of {\em highest
dimension\/} for each eigenvalue $\lambda_1,\ \lambda_2,\, \cdots
\,,\lambda_r,$ respectively. We shall denote the minimal
polynomial through a third list $\| m_1\, m_2\, \cdots\, m_r\,\|$,
called as the type of the minimal polynomial.

Clearly the characteristic~(\ref{PC}) and the
minimal~(\ref{PoliMin}) polynomials of $R^a_{\ b}$ as well as the
eigenvalues are built from the metric and its
derivatives~\cite{PaivaReboucasHallMacCallum1998}. Since they are
either scalars or tensors built from the metric and its
derivatives, the hereditary property~(\ref{PropH0}) can be applied
to them.

Thus, limiting diagrams for the types of the five degree
characteristic polynomial corresponding to the eigenvalue
problem~(\ref{eigeneq}), and for the types of the minimal
polynomial of $R^a_{\ b}$ can be constructed using the hereditary
property~(\ref{PropH0}). For more details about this point see
~\cite{PaivaReboucasTeixeira1997}. From each of these limiting
diagrams (see figures~1 and 2 of~\cite{PaivaReboucasTeixeira1997})
one can construct two distinct diagrams for the limits of Segre
type of $R^a_{\ b}$ in $5$--D. Collecting together the information
of these limiting diagrams one finally obtains the limiting
diagram for Segre types in $5$--D, shown in
figure~\ref{SegreEsp1}.

%%%%%%%%%%%%%%%%%%%%%%%%%%%%%%%%%%%%%%
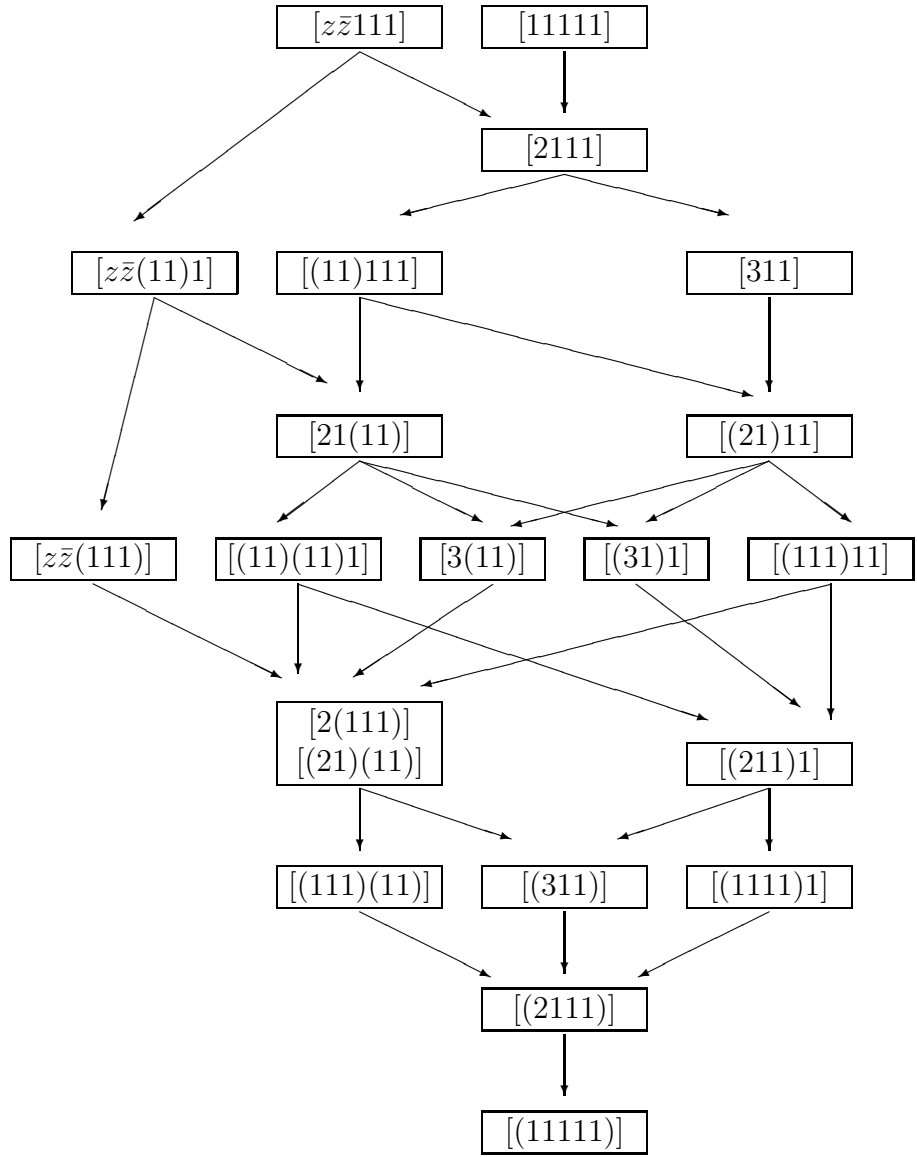
\begin{figure}
\setlength{\unitlength}{3ex}
\begin{center}
\begin{picture}(20,27)(-9.5,-27)

\put(-5,0){\framebox(4,1){[$z\bar{z}$111]}}
\put(0,0){\framebox(4,1){[11111]}}

\put(-3,-0.1){\vector(-4,-3){5.5}}
\put(-3,-0.1){\vector(2,-1){3.2}} \put(2,-0.1){\vector(0,-1){1.5}}

\put(0,-3){\framebox(4,1){[2111]}}

\put(2,-3.1){\vector(-4,-1){4}} \put(2,-3.1){\vector(4,-1){4}}

\put(-10,-6){\framebox(4,1){[$z\bar{z}$(11)1]}}
\put(-5,-6){\framebox(4,1){[(11)111]}}
\put(5,-6){\framebox(4,1){[311]}}

\put(-8,-6.1){\vector(-1,-4){1.3}}
\put(-8,-6.1){\vector(2,-1){4.2}}

\put(-3,-6.1){\vector(0,-1){2.3}}
\put(-3,-6.1){\vector(4,-1){9.6}}

\put(7,-6.1){\vector(0,-1){2.3}}

\put(-5,-10){\framebox(4,1){[21(11)]}}
\put(5,-10){\framebox(4,1){[(21)11]}}

\put(-3,-10.1){\vector(-4,-3){2}} \put(-3,-10.1){\vector(2,-1){3}}
\put(-3,-10.1){\vector(4,-1){6.3}}

\put(7,-10.1){\vector(4,-3){2}} \put(7,-10.1){\vector(-2,-1){3}}
\put(7,-10.1){\vector(-4,-1){6.3}}

\put(-11.5,-13){\framebox(4,1){[$z\bar{z}$(111)]}}
\put(-6.5,-13){\framebox(4,1){[(11)(11)1]}}
\put(-1.5,-13){\framebox(3,1){[3(11)]}}
\put(2.5,-13){\framebox(3,1){[(31)1]}}
\put(6.5,-13){\framebox(4,1){[(111)11]}}

\put(-9.5,-13.1){\vector(2,-1){4.6}}

\put(-4.5,-13.1){\vector(0,-1){2.2}}
\put(-4.5,-13.1){\vector(3,-1){10}}

\put(0.25,-13.1){\vector(-3,-2){3.4}}

\put(3.75,-13.1){\vector(4,-3){4}}

\put(8.5,-13.1){\vector(-4,-1){10}}
\put(8.5,-13.1){\vector(0,-1){3.3}}

\put(-5,-18){\framebox(4,2){\shortstack{[2(111)] \\
{[}(21)(11)]}}} \put(5,-18){\framebox(4,1){[(211)1]}}

\put(-3,-18.1){\vector(0,-1){1.5}}
\put(-3,-18.1){\vector(3,-1){3.7}}

\put(7,-18.1){\vector(-3,-1){3.7}}
\put(7,-18.1){\vector(0,-1){1.5}}

\put(-5,-21){\framebox(4,1){[(111)(11)]}}
\put(0,-21){\framebox(4,1){[(311)]}}
\put(5,-21){\framebox(4,1){[(1111)1]}}

\put(-3,-21.1){\vector(2,-1){3.2}}
\put(2,-21.1){\vector(0,-1){1.5}}
\put(7,-21.1){\vector(-2,-1){3.2}}

\put(0,-24){\framebox(4,1){[(2111)]}}

\put(2,-24.1){\vector(0,-1){1.5}}

\put(0,-27){\framebox(4,1){[(11111)]}}

\end{picture}
\end{center}
\caption{Diagram for the limits of the Segre types of $R^a_{\ b}$
in $5$--D Lorentzian spaces.} \label{SegreEsp1}
\end{figure}

In brief, to achieve this limiting diagram for the Segre
classification of a second order symmetric two-tensor in $5$--D we
have essentially used the hereditary property~(\ref{PropH0})
together with the limiting diagrams for the characteristic and
minimal polynomial types, which are presented in%
~\cite{PaivaReboucasTeixeira1997}.

Improvements of the limiting diagram~\ref{SegreEsp1} can still be
tackled. A first would arise by taking into account the character
of the eigenvectors. A second refinement of the limiting
diagram~\ref{SegreEsp1} can be made by devising a criterion for
separating the Segre types [2(111)] and [(21)(11)], which have the
same type for both characteristic and minimal polynomials. A third
improvement might arise if besides the type of the characteristic
and minimal polynomials one considers the values of their roots.

To close this section we note that although the coordinate-free
technique for finding out limits of space-times in
GR~\cite{PaivaReboucasMacCallum1993} have not yet been extended to
$5$--D space-times, the limiting diagram studied in this section
will certainly be applicable to any coordinate-free approach to
possible limits of non-vacuum space-times in $5$--D.

%********************************************************************
\section{Classification in n--D} \label{class-n}
\setcounter{equation}{0}
%********************************************************************

In this section we shall discuss how one can obtain, by induction, the
algebraic classification and the canonical forms of $R_{ab}$ on a
$n$-dimensional ($n>5$) Lorentzian space, with signature $\rm{(}- +\,
\cdots\, +\rm{)}$, from its classification in $5$--D.

By a procedure similar to that used in the theorem~\ref{eigentheo}, it is
easy to show that $R^a_{\ b}$ defined on an $n$-dimensional ($n \geq 5$)
Lorentzian space has at least one real space-like eigenvector. The
existence of this eigenvector can be used to reduce, by induction, the
classification of symmetric two-tensors on $n$-dimensional ($n > 5$)
spaces (and the corresponding canonical forms) to the classification on
$5$-dimensional spaces (and its canonical forms), thus recovering in a
straighforward way the results of~\cite{SRT-GRG/1995}.

Indeed, with this real non-null $n$--D eigenvector (${\bf\bar{v}}$, say)
one can define another normalized eigenvector ${\bf \bar{u} }$, whose
associated eigenvalue $\bar{\eta}$ is the same of ${\bf \bar{v}}$ in quite
the same way ${\bf v}$ and ${\bf u}$ where defined in~(\ref{vectoru}).
Now, one can follow similar steps (and arguments) in $n$--D to those used
in $5$--D to go from (\ref{rabt1}) to~(\ref{rabz}). Now, introducing a
semi-null basis $\cal B$ for the $n$--D space $T_{p}(M)$, consisting of 2
null vectors and $n-2$ spacelike vectors,
\begin{equation}
\label{basis} {\cal B} =\{{\bf l},{\bf m},{\bf x}^{(1)},{\bf
x}^{(2)},\ldots , {\bf x}^{(n-2)}\}\;,
\end{equation}
such that the only non-vanishing inner products are
\begin{equation} \label{inerp}
{\bf l}.{\bf m}={\bf x}^{(1)}.\,{\bf x}^{(1)}={\bf x}^{(2)}.\,{\bf
x}^{(2)}= \ldots ={\bf x}^{(n-2)}.\,{\bf x}^{(n-2)}=1\;,
\end{equation}
and using the classification for $R^a_{\ b}$ in $5$--D obtained in
section~\ref{class}, it follows that the possible Segre types and the
corresponding canonical forms for $R_{ab}$ in $n$ dimensions are given by
\begin{eqnarray}  \hspace{-1.0cm}
\mbox{\bf Segre type} & & \quad\quad\qquad\  \mbox{\bf Canonical form}
\nonumber \\ {[}1,1\ldots 1] &\quad R_{ab}\, = & 2\rho_{1}l_{(a}m_{b)} +
\rho_{2}(l_{a}l_{b} + m_{a}m_{b}) + \rho_{3}x^{(1)}_{a}x^{(1)}_{b} +
\rho_{4}x^{(2)}_{a}x^{(2)}_{b}
 \nonumber  \\
 & & + \cdots + \rho_{n}x^{(n-2)}_{a}x^{(n-2)}_{b}\,, \label{rab11111n}  \\
{[}21\ldots 1] &\quad R_{ab}\, = & 2\rho_{1}l_{(a}m_{b)} \pm l_{a}l_{b} +
\rho_{3}x^{(1)}_{a}x^{(1)}_{b} + \rho_{4}x^{(2)}_{a}x^{(2)}_{b}
 \nonumber  \\
 & & + \cdots + \rho_{n}x^{(n-2)}_{a}x^{(n-2)}_{b}\,, \label{rab2111n}  \\
{[}31\ldots 1] &\quad R_{ab}\, = & 2\rho_{1}l_{(a}m_{b)} + 2l_{(a}x_{b)} +
\rho_{1}x^{(1)}_{a}x^{(1)}_{b} + \rho_{4}x^{(2)}_{a}x^{(2)}_{b}
 \nonumber  \\
 & & + \cdots + \rho_{n}x^{(n-2)}_{a}x^{(n-2)}_{b}\,,  \label{rab311n}  \\
{[}z\bar{z}11\ldots 1] &\quad R_{ab}\, = & 2\rho_{1}l_{(a}m_{b)} +
\rho_{2}(l_{a}l_{b} - m_{a}m_{b}) + \rho_{3}x^{(1)}_{a}x^{(1)}_{b} +
\rho_{4}x^{(2)}_{a}x^{(2)}_{b}
 \nonumber  \\
 & & + \cdots + \rho_{n}x^{(n-2)}_{a}x^{(n-2)}_{b}\,, \label{rabZZn}
\end{eqnarray}
and the degeneracies thereof. Here the coefficients $\rho_{1}, \ldots
,\rho_{n}$ are real scalars and $\rho_{2}\neq 0$ in (\ref{rabZZn}).

This classification of symmetric two-tensors in any $n$-dimensional spaces
and their canonical forms are important in the context of $n$-dimensional
brane-worlds as well as in the framework of $11$--D supergravity and
$10$--D superstrings.

%********************************************************************
\section{Further Results}  \label{FurtherRes}
\setcounter{equation}{0}
%********************************************************************

In this section we shall briefly discuss some recent
results~\cite{Note} concerning the algebraic structure of a second
order symmetric tensor $R$ defined on a $5$--D Lorentzian manifold
$M$, which can be collected together in the following theorems:

\vspace{2mm}
\begin{theorem} \label{rabtheo1}
Let $M$ be a real $5$-dimensional manifold endowed with a Lorentzian
metric $g$ of signature ${\rm(} - + + + + {\rm)}$. Let $R^a_{\ b}$ be the
mixed form of a second order symmetric tensor $R$ defined at a point $p
\in M$. Then
\begin{enumerate}
\item
$R^a_{\ b}$ has a time-like eigenvector if and only if it is
diagonalizable  over $\R$ at $p$.
\item
$R^a_{\ b}$ has at least three real orthogonal independent
eigenvectors at $p$, two of which (at least) are space-like.
\item
$R^a_{\ b}$ has all eigenvalues real at $p$ and is not diagonalizable if
and only if it has an unique null eigendirection at $p$.
\item
If $R^a_{\ b}$ has two linearly independent null eigenvectors at $p$ then
it is diagonalizable over $\R$ at $p$ and the corresponding eigenvalues
are real.
\end{enumerate}
\end{theorem}

Before stating the next theorem we recall that the $r$-dimensional
($r \geq 2$) subspaces of $T_{p}(M)$ can be classified according
as they contain more than one, exactly one, or no null independent
vectors, and they are respectively called time-like, null and
space-like $r$-subspaces of $T_p(M)$. Space-like, null and
time-like $r$-subspaces contain, respectively, only space-like
vectors, null and space-like vectors, or all types of vectors.

\vspace{2mm}
\begin{theorem} \label{rabtheo2}
Let $M$ be a real $5$-dimensional manifold endowed with a Lorentzian
metric $g$ of signature ${\rm(} - + + + + {\rm)}$. Let $R^a_{\ b}$ be the
mixed form of a second order symmetric tensor $R$ defined at a point $p
\in M$. Then
\begin{enumerate}
\item
There always exists a 2--D space-like subspace of $T_p(M)$
invariant under $R^a_{\ b}$.
\item
If a non-null subspace $\cal V$ of $T_p(M)$ is invariant under
$R^a_{\ b}$, then so is its  orthogonal complement ${\cal V}^{\perp}$.
\item
There always exists a 3--D time-like subspace of $T_p(M)$
invariant under $R^a_{\ b}$.
\item
$R^a_{\ b}$ admits a $r$-dimensional {\rm(}$r=2, 3,4${\rm )} null
invariant subspace $\cal N$ of $T_p(M)$ if and only if $R^a_{\ b}$
has a null eigenvector, which lies in $\cal N$.
\end{enumerate}
\end{theorem}

The proofs can be gathered essentially from the canonical forms
(\ref{rab11111})~--~(\ref{rabzz111}) and are not presented here
for the sake of brevity, but the reader can find them in details
in reference~\cite{Note}.

%********************************************************************
\section{Concluding Remarks}  \label{ConcluRem}
\setcounter{equation}{0}
%********************************************************************

Recent results on the expansion of the bulk geometry have
raised some experimental and theoretical difficulties to $5$--D
brane-world models (for a clear indication of these problems
see Damour \emph{et al.\/}~\cite{Damour}).
In particular, the massive gravitons (scalar-like polarization
states) coupling to matter modify the usual general relativistic
relation for interaction of matter and light, giving rise to
a sharp discrepancy in the bending of light rays by the Sun, for
example, so well explained by general relativity. Furthermore,
in a recent paper~\cite{Durrer} Durrer and Kocian have obtained a
modification of Einstein quadrupole formula for the emission of
gravity waves by a binary pulsar in the framework of $5$--D
brane-worlds. They have computed the induced change for the binary
pulsar PSR 1913+16 and shown that it amounts to about 20\% in sharp
contradiction with the current observations. Such observational
problems challenge the brane-world scheme --- a deeper
understanding of the nature of the bulk, its dynamics, and how the
embedding relates the bulk geometry to that of the brane, clearly
is necessary.

The Segre classification, together with the limiting diagram in
$5$--D, and the extension of Geroch's limit could, in principle,
offer ways of circumventing the current serious difficulties
in $5$--D brane-world. In fact, on the one hand it is well know
that the Riemann tensor is decomposable into three irreducible parts,
namely the Weyl tensor, the Ricci tensor and the Ricci scalar.
On the other hand, it is known that the embedding is defined by
the components of the Riemann tensor of the bulk (see, e.g.,%
~\cite{Monte,Maia}). Therefore one can potentially use the algebraic
classification of the Weyl and Ricci parts of the $5$--D curvature
tensor to shed light into the above mentioned problem faced by $5$--D
brane-world models. So, for example, the embedding conditions can
in principle be used to relate Segre types in the $5$--D bulk with
Segre types of the $4$--D brane-world (we note the limiting diagrams
and Geroch's limiting theorem of the present paper hold for a fixed $n$).
One could also find out, e.g., for which Segre and Petrov types (if any)
associated to Riemann tensor of the bulk the above mentioned problems
would not come about. Such an interesting research project does not
seem to be straightforward, and is beyond the scope of the
present review, though.
We note in passing that this project would require not only the Segre
classification but also the Petrov types of the Weyl tensor in
$5$--D, and to the best of our knowledge, the classification of
Weyl tensor in five and higher dimensions has not yet been performed
(for the $5$--D case note, however, the section~2 of the recent article 
by De Smet~\cite{Smet}).
We also emphasize that although recent interest in $5$--D brane-world
models has motivated this paper, our results apply to any second
order symmetric tensor in the context of any locally Lorentzian 
geometrical theory.

To close this article we mention that it has sometimes been assumed, in
the framework of brane-world cosmologies, that the source term (${\cal
T}_{ab}$ in $4$--D) restricted to the brane is
a mixture of ordinary matter and a minimally coupled scalar field%
~\cite{HogenColeyHe2002,HoogenIbanez2002}, where the gradient of the
scalar field $\phi_\alpha \equiv \phi_{;\,\alpha}$ is a time-like vector.
In these cases the scalar field can mimic a perfect fluid, i.e., it is of
Segre type $[1,(111)]$. However, according to ref.~\cite{SanRebTei},
depending on the character of the gradient of the scalar field (and also
on the character of the gradient of the corresponding potential) it can
also mimic : (i) a null electromagnetic field and pure radiation (Segre type
$[(211)]$, when $\phi^\alpha$ is a null vector); (ii) a tachyon fluid
(Segre type $[(1,11)1]$, when $\phi^\alpha$ is a space-like vector); and
clearly (iii) a $\Lambda$ term (Segre type $[(1,111)]$, when the scalar
field is a constant).

\section*{Acknowledgments}
M.J. Rebou\c cas thanks J.A.S. Lima for his kind hospitality at
Physics Department of Federal University of Rio Grande do Norte
(DFTE-UFRN). He also acknowledges CNPq for the grant under which
this work was carried out.

\end{document}